%% file: main.tex
\documentclass[titlepage,twocolumn,hidelinks,10pt,byrevtex,secnumarabic,nofootinbib,longbibliography,a4paper,showkeys,preprintnumbers]{revtex4-2}
\usepackage{graphicx}
\usepackage{amsmath}
\usepackage{amssymb}
\usepackage{float}
\usepackage{censor}
\usepackage{moreverb}
\usepackage{url}
\usepackage{tabularx,array}

\usepackage{xr-hyper}
\usepackage{xcolor}
\usepackage{colortbl}
\usepackage[final,commandnameprefix=always]{changes}
\usepackage{tikz}
\usepackage{pgfplots}
\usepackage{listings}
\usepackage{datetime2}
\DTMsetstyle{iso}
\DTMsetup{showseconds=false,showzoneminutes=false}

\input{zdoi.tex}

\newcommand{\zenodo}[1]{\doi{10.5281/zenodo.#1}}
\usetikzlibrary{shapes,arrows,arrows.meta,positioning,calc,external,patterns}
\usepgfplotslibrary{fillbetween}
\pgfplotsset{compat=1.11}

\usepackage[pdftex,
pdfauthor={Jose-Maria Martin-Olalla},
pdftitle={Characterizing the Carnot cycles at absolute zero},
            pdfsubject={Foundations of Thermodynamics},
            pdfkeywords={second law of thermodynamics; third law of thermodynamics; specific heat; statistical mechanics; entropy; temperature; Carnot engine; foundations of thermodynamics; Carnot theorem; absolute zero},
            pdfproducer={Latex with hyperref, or other system},
            pdfcreator={pdflatex, or other tool}]{hyperref}

                        \newcommand\rorlink[2]{\href{https://ror.org/#2}{#1\,\includegraphics[scale=0.1]{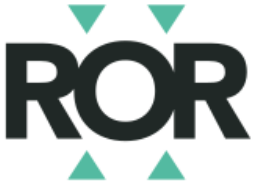}}
                        }

\usepackage{orcidlink}
\providecommand{\orcidlinki}[2]{\href{https://orcid.org/#2}{#1}\orcidlink{#2}}
            \usepackage{breakurl}
\usepackage{siunitx}
\usepackage{mathtools}
\usepackage[utf8]{inputenc}
\usepackage[english]{babel}
\usepackage{lineno}
\usepackage{array}

\usepackage{ccicons}

\AtBeginDocument{\usepackage{booktabs}}

\makeatletter\AtBeginDocument{\let\@elt\relax}\makeatother

\newcommand{\uaddress}{\rorlink{Universidad de Sevilla}{03yxnpp24}, Facultad de Física, Departamento de Física de la Materia Condensada,  ES41012 Sevilla, Spain}

\begin{document}
\author{\orcidlinki{José-María Martín-Olalla}{0000-0002-3750-9113}}
\affiliation{\uaddress}
\email{olalla@us.es}
\thanks{\ccby}

\homepage{https://twitter.com/MartinOlalla\_JM}

\title{Characterizing the Carnot cycle at absolute zero: a reply to ``Comment on `Proof of the Nernst theorem' '' }

\received[Submitted ]{March 4, 2026}

\revised{May 22, 2026}
\accepted{July 19, 2026}
\published{August 3, 2026}

\preprint{\textcolor{blue}{This is the Author's accepted manuscript. Existe una versión en lengua castellana.  \copyright The Author, 2026}}
\preprint{\textcolor{blue}{The Version of Record is published by \emph{The European Physical Journal Plus} (2026) \textbf{141} 900 \doi{10.1140/epjp/s13360-026-08111-8}}}
\begin{abstract}
  \input{abstract.tex}

\textcolor{blue}{The Version of Record is published by \emph{The European Physical Journal Plus} (2026) \textbf{141} 900 \doi{10.1140/epjp/s13360-026-08111-8}}
  
    This version is also published in \texttt{Zenodo} \zenodo{\zdoi} and at the Institutional repository \burl{https://idus.us.es/handle/11441/188827}.

    Total word count: \texttt{\input{mainCount.tex}}

\end{abstract}

\keywords{\input{keywords.tex}}

\maketitle

\input{manuscript.tex}

\acknowledgments
The author acknowledges the use of an Artificial Intelligence large language model, Gemini, to refine the draft manuscript for grammar and clarity. Gemini was made available to the author through a collaborative initiative between Google and Universidad de Sevilla.

The author thanks the staff at the Library and the Communications Service of the Universidad de Sevilla for their assistance in archiving and promoting the original manuscript.

\chadded{The visualization in Figure~\ref{fig:carnot} and the accompanying animation (Supplementary Material) were already being prepared for the VII Congreso Nacional de Estudiantes de Física (Conef \textsc{VII}), Universidad de Extremadura (RoR \href{https://ror.org/0174shg90}{0174shg90}), Badajoz (Spain), March 5--7, 2026. The author thanks the Board of Conef \textsc{VII} for their kind invitation and hospitality, and remains grateful to the attendees.}

Funding for open access publishing: Universidad de Sevilla/CBUA. Open Access provided thanks to CRUE-CSIC agreement with Springer Nature.

\input{main.bbl}
\input{history.tex}

\end{document}

%% file: zdoi.tex
\newcommand{\zdoi}{21545290}

%% file: abstract.tex
This reply addresses a recent comment concerning the proof of the Nernst theorem. I clarify how a Carnot engine can consistently operate at $T=0$ through a continuous deformation of a cycle operating at $T>0$. By examining the limit where heat exchange with the cold reservoir vanishes, I show that the Nernst theorem ensures that the concept of temperature remains physically consistent at the absolute zero limit.

%% file: mainCount.tex
1812

%% file: keywords.tex
Nernst theorem; Einstein-Nernst debate

%% file: manuscript.tex
I appreciate the recent comment by \citet{Chen2026} regarding my study on the proof of the Nernst theorem \cite{Martin-Olalla2025c}. I concur with the Authors' assertion that it is of the utmost importance to clarify the relationship ``between the Carnot cycle and the fundamental laws of thermodynamics.'' I offer this reply to clarify why the Nernst theorem \cite{Nernst-1924}:
\begin{equation}
  \label{eq:1}
  \lim_{T\to0} (S(T,x_2)-S(T,x_1))=0,\quad \forall x_2,x_1\in\mathcal{D},
\end{equation}
where $x$ is a mechanical variable with domain $\mathcal{D}\subset\mathbb{R}^+$, is directly derived from the statement of the Second Law. To put it simply: if the temperature $T$ in Equation~(\ref{eq:1}) is the Carnot temperature, and the entropy $S$ is the Clausius entropy, then the theorem is a formal consequence of the Second Law \cite{Martin-Olalla2025c}.

With $S(T,x)$ bounded in the neighborhood of $T=0$ as dictated by stability conditions \cite{Bazarov1971,Martin-Olalla2025d,Martin-Olalla2026d}, evaluating the limit in Equation~(\ref{eq:1}) is fundamentally a physical question rather than a complex mathematical issue. The left-hand side represents the isothermal change of entropy, $\Delta S$, during a reversible change in $x$, which is given by $\mathsf{Q}/T$, where $\mathsf{Q}$ is the heat exchanged during the process. Classical thermodynamics prescribes no a priori sign or value for $\Delta S$; it can be positive (heat absorbed), negative (heat rejected), or zero (no heat required to change $x$). The Nernst theorem asserts that $\Delta S$ must shrink to zero universally as $T$ approaches zero. Physically, this means that at $T=0$, distinct thermodynamic states do not differ in entropy despite having different macroscopic configurations, as exemplified by the solid-liquid phase equilibrium of helium near absolute zero.

The proof presented earlier\cite{Martin-Olalla2025c} is based on the idea that $T=0$ must be determined by Carnot's theorem given that $T$ is Carnot's temperature. \citet{Chen2026} claim that the Carnot's theorem ``is only suitable for the temperature region of $T>0$.'' This aligns with \citet{Einstein-1913}'s perspective at the Second Solvay Congress: by denying that a Carnot engine can operate ``in practice'' at $T=0$, the Carnot theorem is deemed inapplicable at this limit, consequently restricting the temperature domain to $T>0$.

In my view, this restriction is inconsistent with the framework of the Second Law, which establishes temperature as a primary consequence and provides its metric via the Carnot theorem. We do not restrict the domain of length to $L>0$ by excluding $L=0$; we do not restrict mass to $m>0$; nor do we restrict the magnitude of heat exchanges to $Q>0$. Consequently, we should not restrict the domain of temperature to $T>0$. Instead, we should derive the implications of a null reversible heat exchange with the cold reservoir—the only consistent way to achieve $T=0$ within the Second Law. That consequence is precisely the Nernst theorem via the proof presented earlier.\cite{Martin-Olalla2025c}

This approach requires a conceptual characterization of a null measurement of $T$ using a Carnot engine. In the Discussion of the original study, I sketched such a procedure but certainly lacked sufficient clarity. I now offer a more detailed description based on a previous study \cite{Martin-Olalla2003b}.

\begin{figure*}
  \centering
  \includegraphics{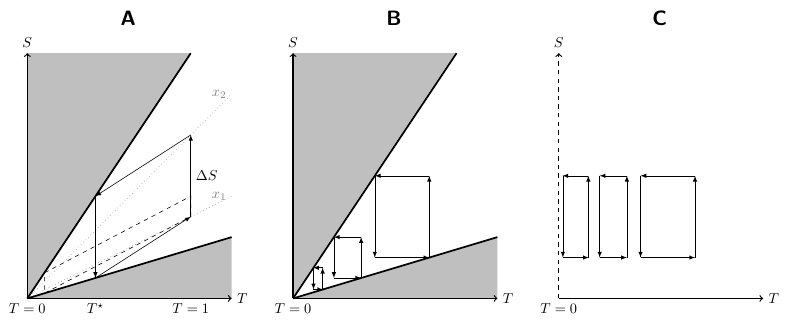}
  \caption{\textbf{A.} A substance compliant with Nernst's theorem and a Carnot cycle (black solid lines with arrows) where the two isotherms are connected by two arbitrary paths with a constant entropy shift $\Delta S$, maintaining the net balances of a two-isotherm, two-adiabat Carnot cycle; specifically the area of the cycle remains $\Delta S\times\Delta T$.  The gray areas display non-accessible pairs of $(S,T)$ within the domain $\mathcal{D}\subset\mathbb{R}$ of $x$. The temperature of the low isotherm $T^\star$ is determined by equaling the full range of entropy within the domain of $x$ to the entropy $\Delta S$ exchanged at the high isotherm. By shrinking $\Delta S$ at the high isotherm, $T^\star$ continuously decreases, exemplified by the cycle in dash stroke. The gray dotted line shows iso-$x$ lines, with positive slope as required by stability. An animation of this plot is available as supplementary information. \textbf{B.} The same substance and standard two-isotherm, two-adiabat Carnot engines. They coalesce into a single point in the limit $T \to 0$ because both $\Delta T$ and $\Delta S$ must vanish; consequently, this cycle cannot operate at $T=0$ \cite{liboff-physicsessays-94}. \textbf{C.} For a hypothetical substance that violates the Nernst theorem, the Carnot engine could theoretically be placed anywhere on the $TS$ plane, except at $T=0$ where the cycle is deemed unpractical. The broken line at $T=0$ highlights that reversible process in $T=0$ are excluded in practice.\cite{Einstein-1913} Only configuration A allows for a consistent thermodynamic characterization of $T=0$.}
  \label{fig:carnot}
\end{figure*}

The Carnot engine is a fundamental model because it is the simplest engine compliant with Planck’s statement of the Second Law: an engine with only two net heat exchanges. While this is typically achieved via two reversible isotherms and two reversible adiabats---rendered in a $TS$ diagram by the well-known rectangular cycle with an area equal to $\Delta T \times \Delta S$---, it can also be realized through a cycle composed of two reversible isotherms exchanging entropy $\Delta S$ with high and low reservoirs, connected by two arbitrary paths that differ by a constant entropy shift equal to $\Delta S$ (see Figure~\ref{fig:carnot}A). In this generalized construction, the heat and entropy exchanges along these arbitrary paths cancel out locally---i.e., at any given $T$---and thus globally, ensuring that the net energy and entropy balances are governed by the same equations as a standard Carnot cycle: simple geometry shows that the area enclosed by the cycle remains $\Delta T \times \Delta S$.

At the high isotherm, $\Delta S$ is extracted by reversibly changing the mechanical parameter from $x_1$ to $x_2$. For the low isotherm, the Nernst theorem implies the existence of $T^\star$ such that $\Delta S^\star(T^\star,\mathcal{D})=\Delta S$, where $\Delta S^\star(T,\mathcal{D})$ is the change in entropy when the mechanical variable is extended across its entire domain $\mathcal{D}$ (e.g., from zero to infinite magnetic field for a paramagnetic substance). With $\Delta S$ given, the working substance cannot perform this cycle at any temperature below $T^\star$.

To decrease $T^\star$, an experimentalist needs only reduce the change in $x$ at the unit (high) isotherm so that $\Delta S$ decreases. However, at the low isotherm, the change in $x$ continues to span the domain $\mathcal{D}$. As $\Delta S$ shrinks toward zero, $x_2$ continuously approaches $x_1$ at the high isotherm. At the low isotherm, the entropy exchange also shrinks to zero, yet the change in the mechanical variable continues to span the domain $\mathcal{D}$. In the limit $T=0$, the four-stroke cycle degenerates into a reversible two-stroke path. Notably, at $T=0$, no reversible isothermal change of $x$ is required; the ``cycle'' simply contacts the isotherms at a single value of $x$. This degeneracy does not invalidate the construction; rather, it is expected for a null measurement ---much like how the jaws of a caliper meet when measuring zero length. \citet{Einstein-1913}'s remark no longer applies because, in the limit, no change is attained at absolute zero as shown in Ref.~\cite[Figure~1]{Martin-Olalla2025c}. Importantly, the cycle operating at $T=0$ is not a ``reversible adiabatic cycle'' as \citet{Chen2026} suggest, but a reversible diathermal path whose net heat and entropy balances are zero.

I must note that this construction is not possible with the standard Carnot engines (two isotherms and two adiabats) described by \citet{Chen2026} and neither is possible when the Nernst theorem is negated and \citet{Einstein-1913}'s remark accepted. For the standard engines, the limiting temperature is determined by the intersection of the reversible adiabatic path (a horizontal iso-$S$ line) and the boundary of accessible states. In that framework, the experimentalist can never exhaust the domain of $x$ at the low isotherm. Furthermore, as $T\to0$, the temperature of the high isotherm cannot remain fixed, the extent of the reversible adiabatic path also shrinks, and the cycle eventually coalesces into a single point, instead of a finite, reversible, two-stroke, path, see figure~\ref{fig:carnot}B. This explains why, after assuming the Nernst theorem, \citet{liboff-physicsessays-94} concluded that at $T=0$ ``a Carnot engine does not exist and the Kelvin temperature scale is not defined,'' a conclusion clearly inconsistent with the fact that temperature is a fundamental physical observable.

Finally, if the Nernst theorem is negated while Planck’s statement is upheld, a similar construction fails. While one can move the low isotherm of a Carnot engine toward zero so that the heat exchange shrinks, see figure~\ref{fig:carnot}C, if the entropy exchanged remains non-zero, then at $T=0$ the engine ceases to exist ``in practice,'' as per \citet{Einstein-1913} argument. No null measurement is achieved; it is simply excluded---this would be akin to a caliper disappearing the moment its jaws touch. It is this abrupt exclusion that is formally inconsistent with continuity and brings a series havocs \cite{Nernst-1924,Martin-Olalla2026}. In the published study and in this reply I only highlighted that $T=0$ remains formally uncharacterized if the Nernst theorem is negated and \citet{Einstein-1913}'s argument accepted. This is paradoxical for a physical observable deduced from a law of nature and is inconsistent the moment $T\to0^+$ is present in the statement of the theorem. If the $T$ in the theorem is Carnot's temperature, the only temperature deduced from Planck's statement, then the theorem is proven.

For completeness, the fact that $S(U,x)$ is concave and its slope is infinite at $T=0$ does not imply that entropy is unbounded. For instance, $S=\sqrt{U}$ is concave and finite as $\partial S/\partial U\to\infty$. The relationship between stability and thermodynamic properties at $T\to0^+$ is best understood by analyzing $U(S,x)$, which is always convex, and is flat at $T=0$. Given $U_{ss}=(\partial^2 U/\partial S^2)_x=T/C_x$, stability $U_{ss}>0$ at $T=0$ mandates that $C_x$ vanishes at least as fast as $T$.\cite{Bazarov1971,Martin-Olalla2025d} This implies $U(S,x)$ is parabolic or subparabolic, ensuring the horizontal tangent at $T=0$ is attained at a finite value of $S$.

%% file: history.tex
\section*{Timeline}
\label{sec:timeline}

The author received an email from the Editorial Office on March 3rd, 2026 informing him that the Comment by \citet{Chen2026} had been accepted and inviting the Author to reply. The invitation was accepted after a reading of the Comment, and the reply was drafted and submitted on March 4th, 2026.

These events were coincidental with the drafting of the talk scheduled for March 6th 2026 at Conef \textsc{VII}.

 The writing of the manuscript required \texttt{\input{versionCount.tex}} different \texttt{emacs} sessions from \texttt{\input{startDate.tex}} to \mbox{\texttt{\input{endDate.tex}}.}

%% file: versionCount.tex
17

%% file: startDate.tex
2026-03-03 23:16

%% file: endDate.tex
2026-08-03 23:14